# Adaptive multi-wave sampling for efficient chart validation

Georg Hahn, Sebastian Schneeweiss, Shirley Wang

Brigham and Women's Hospital, Division of Pharmacoepidemiology and Pharmacoeconomics, and Department of Medicine, Harvard Medical School, Boston, MA 02120, USA

**Abstract**

Computable phenotypes are used to characterize patients and identify outcomes in studies conducted using healthcare claims and electronic health record data. Chart review studies establish reference labels against which computable phenotypes are compared to understand their measurement characteristics, the quantity of interest, for instance the positive predictive value. We describe a method to adaptively evaluate a quantity of interest over sequential samples of charts, with the goal to minimize the number of charts reviewed. With the help of a simultaneous confidence band, we stop the reviewing once the confidence band meets a pre-specified stopping threshold. The contribution of this article is threefold. First, we tested the use of an adaptive approach called Neyman's sampling of charts versus random or stratified random sampling. Second, we propose frequentist confidence bands and Bayesian credible intervals to sequentially evaluate the quantity of interest. Third, we propose a tool to predict the stopping time (defined as the number of charts reviewed) at which the chart review would be complete. We observe that Bayesian credible intervals proved to be tighter than its frequentist confidence band counterparts. Moreover, we observe that simple random sampling is often performing similarly to Neyman's sampling.

**Introduction**

There is increasing use of claims and electronic health record databases to evaluate the safety and effectiveness of medical products[1,2,3]. For such studies, computable phenotypes are used to measure patient characteristics and outcomes. However, their measurement is subject to error; minimizing such error and quantitatively assessing the arising bias is essential for causal inference[4,5]. Validation studies are needed to understand the measurement characteristics of computational phenotypes compared to medical charts as an accepted reference, for instance their specificity or positive predictive value[6]. The core of a validation study is a set of labeled charts that provide a reference standard. However, establishing such labeled charts is expensive and time consuming as medical experts need to review and label each chart. Approaches to minimize the number charts that need to be labeled is desirable[8,9].

Previous work considered a Bayesian framework for a validation study design which allows one to detect when sufficient validation data for a binary measure has been collected, whereby charts were drawn in random batches[7]. Newer work employs Neyman sampling and raking weights assuming a fixed number of charts to review and samples to spend[10]. The approach is not data adaptive and the authors did not compare it to alternative sampling approaches. Other work considered several sampling strategies including Neyman's sampling in a classical measurement-error framework, but in a parametric model context and without the use of confidence bands or Bayesian credible intervals[11]. To the best of our knowledge, the focus of our work, precisely the evaluation of adaptive multi-wave Neyman's sampling, has not been previously considered.

In this work we considered the use of an efficient sampling process in which batches of charts are adaptively sampled and evaluated sequentially to decide if a parameter of interest, such as the PPV, exceeds one (or more) pre-specified stopping thresholds that would indicate sufficiently high performance or precision of the algorithm that further validation is unnecessary. Similarly, if the sequential evaluations indicate futility, in other words, that with infinite additional samples, we do not expect to observe desirable performance for a particular algorithm, then spending additional resources on chart review can be avoided. Furthermore, an adaptive sampling process for chart validation can be used to track and predict time dependent quantities. For example, it may be of interest to predict when a validation study will meet a pre-specified stopping threshold to aid in planning a validation study, assuming this cost is proportional to the number of charts involved in the validation study.

Therefore, we present an adaptive, multi-wave sampling validation algorithm for the scenario in which we have data consisting of a (binary) outcome variable that is being validated and several patient characteristics that may or may not be risk factors for the outcome variable. For illustrative purposes, the main quantity of interest from the chart validation study is the PPV for the proposed algorithm given the chart review reference standard. We select two thresholds to encode a stopping criterion, for example, once we are confident that the lower bound of the sequentially evaluated PPV lies above some pre-specified threshold $\tau_1$, or that the upper bound of the PPV lies below some pre-specified threshold $\tau_2$ (for instance $\tau_1=\tau_2=0.8$).

We are interested in exploring three aspects of the validation. First, we explore four strategies for sampling charts, called random sampling (sampling which disregards baseline covariates), two variants of stratified random sampling (equal-size batches across strata defined by baseline covariates, or proportional to the strata sizes), and Neyman's sampling. The latter adapts the sample size selected based on the variance of the outcome in each stratum. Second, we explore the performance of our validation process when using two different approaches to compute confidence bands around the sequentially evaluated PPV. These are simultaneous confidence bands of Lai[12], and Bayesian credible intervals. Third, we investigate the issue of forecasting when the stopping criterion will be met while the study is running, in the sense that at any point in time, we compute an estimate of the remaining number of chart samples to review.

The algorithm presented in this paper has been implemented in the R-package "chartreview", available on CRAN[13].

**Methods**

This section first introduces the types of sampling we consider. Those are random sampling, stratified random sampling (equal-size batches across strata, or proportional to the strata sizes), and Neyman's sampling. We then define the types of confidence bands we use, in particular confidence bands and Bayesian credible intervals for a binomial quantity, and normal intervals for a continuous quantity. Moreover, we add a few remarks about raking. The section concludes with an algorithm to perform the chart validation.

*Sampling strategies*

We are given data for $N \in \mathbb{N}$ individuals, encoded in a matrix $X \in \mathbb{R}^{N*p}$ and a vector $y \in \mathbb{R}^N$. We assume that in each row, X contains data on $p \in \mathbb{N}$ patient baseline characteristics which may or may not be risk factors associated with y, while y contains some response (outcome) of interest that is measured with an unknown degree of error. Moreover, we assume that the data in X can be stratified in $m \in \mathbb{N}$ strata, which will determine how the sampling process is carried out. We are interested in computing some statistic T of interest of y, for instance the PPV.

To adaptively estimate T(y), we assume we can iteratively draw samples from y according to some sampling strategy. The sampling of any stratum will stop as soon as it is depleted, while the other non-depleted strata may continue to be sampled. The number of batches depends on both the reservoir of samples as well as on the batch size $B \in \mathbb{N}$ we employ for sampling. Each time we draw samples from X, we draw samples for the same individuals from y, which are then used to approximate T(y).

Four sampling rules are considered in this article in order to draw samples from X and y, which are defined as follows:

1. We draw B samples per wave while disregarding any strata information. This will be referred to as "random sampling".
2. We draw [B/m] samples per stratum, where the brackets denote the floor operation that rounds down to the nearest integer. This will be referred to as "stratified random sampling 1" or "stratified1".
3. We draw [B*$w_s$] samples for each stratum s∈{1,...,m}, where $w_s$∈[0,1] is the proportion of samples in stratum s with respect to the total number of samples. This will be referred to as "stratified random sampling 2" or "stratified2". Note that for this method, we always weigh with respect to the initial proportion of samples in the strata, as opposed to the proportions while sampling is in progress.
4. The fourth technique, called "Neyman's sampling", works as follows (Shepherd et al., 2021). Denote with $N_s$ the sample size in each stratum s∈{1,...,m} with respect to the data that has already been validated. We aim to compute the new allocation of samples to each stratum in wave k. Denote with $n_i$ the overall number of samples spent in each wave i (in our case, this will be the batch size). Then,

$$\sum_{i=1}^{k} n_i$$

is the total number of samples spent in the first k waves. Denote the standard deviation of the data in each stratum s∈{1,...,m} as $\sigma_s$>0 (with respect to the data that has already been validated). Neyman's sampling computes the new sample size $n_{k,s}$ for the current wave k and for each stratum s∈{1,...,m} as

$$n_{k,s} = \left(\sum_{i=1}^{k} n_i\right) \frac{N_s \sigma_s}{\sum_{t=1}^{m} N_t \sigma_t} - \left(\sum_{i=1}^{k-1} n_{i,s}\right),$$

where the sample sizes ($n_{k,1}$,...,$n_{k,m}$) can additionally be reweighted to sum up to the batch size B in order to spend exactly B new samples in each wave. The goal behind Neyman's sampling is to optimize sample allocation (with respect to minimizing the variance) based on the variance within strata.

Three computational aspects are of importance. First, the standard deviation per stratum needed for Neyman's sampling can be volatile at the start of the sampling process when sample sizes are low. Therefore, a robust estimator of the variance such as the median absolute deviation (mad) can be more suitable than the conventional estimator.

Second, the batch size plays an important role at the start of the sampling since larger batch sizes at the start likewise reduce volatility in the estimates. In later waves, when each stratum has already been sampled from and the standard deviation estimate is more stable, the choice of the batch size is less important.

Third, in Neyman's sampling it can happen that the sample size of a stratum for the new wave is reduced to zero. In this case, no further sample will be drawn in future waves, which might introduce bias. It can therefore be advantageous to introduce a minimal batch size per stratum to prevent the scenario in which a stratum is excluded from being sampled in all future waves.

*Confidence bands*

After having drawn a new batch of B samples from y, we aim to update our knowledge on the quantity T(y) which we aim to estimate. To this end, we aim to compute valid confidence bounds for T(y).

In the case that T is the PPV, and y is binary, we compute binomial confidence bounds for T(y). We consider two options. The first are the confidence bands of Lai[12], and the other are Bayesian credible intervals.

For Lai's confidence bands, we assume that in any wave, we have observed $s \in \mathbb{N}$ successes among $k \in \mathbb{N}$ samples being drawn from y. The overall error probability is denoted by $\alpha \in (0,1)$. To compute Lai's bounds, we solve $(k+1)b(k,p,s)=\alpha$ for p, where $b(k,p,s)=k!/(s!(k-s)!)\, p^s\,(1-p)^{k-s}$ is the density of the binomial distribution. The aforementioned equation will have two solutions $p_1$ and $p_2$ for fixed α, which then form the confidence interval $[p_1,p_2]$ for p. As proven in Lai (1976), this construction yields valid confidence bounds with an overall coverage of 1-α which can be updated as more samples are drawn and thus more successes are observed.

To compute Bayesian credible intervals, we consider a Beta-Binomial model, given by a Beta(1,1) prior on the unobserved parameter p which is then updated once Binomial samples are observed. This results in a Beta(1+s,1+k-s) posterior after having observed s successes among k samples. To arrive at a credible interval, we compute the α/2-quantile $q_{\alpha/2}$ and the (1-α/2)-quantile $q_{1-\alpha/2}$ of the Beta(1+s,1+k-s) posterior, which then form the credible interval $[q_{\alpha/2},q_{1-\alpha/2}]$. Note that no alpha spending is required in the Bayesian case.

The previous approach can be extended in a straightforward fashion to a continuous quantity being validated. In this case, we compute normal confidence bands. Further details are provided in the Appendix.

*Raking weights*

When sequentially evaluating T(y), we aim to ensure that the distribution of key risk factors in the binary samples drawn up to each wave resembles the distribution observed in the overall population. We correct for finite sample behavior with the help of a procedure called raking. Raking allows us to first identify if the sampled population distribution of risk factors is discrepant from the overall population, and second compute raking weights to correct it. We apply the R-package "anesrake" (available on CRAN[14]) to the vector y.

*Adaptive multi-wave chart review sampling process*

The complete adaptive chart review sampling process is shown in Figure 1. In a prototypical setting, the investigators would create a chart review protocol or annotation guide for the phenotype or outcome of interest. A claims based algorithm for the outcome would be proposed and used to classify patients as either having or not having the endpoint of interest. Sequential samples of patients can then be identified for multi-wave chart review. These samples can be obtained via random sampling, stratified random sampling 1 and 2, or Neyman's sampling. Each sample will contain data in X and y, where X contains data on patient risk factors that define strata for stratified sampling, and y reflects the error-prone claims based algorithm classification of the endpoint. The algorithm works along the following steps.

First, the data are sampled. We use either random sampling, stratified random sampling 1 and 2, or Neyman's sampling to determine how to allocate a new batch of B samples to the m strata defined by x. Then trained chart reviewers evaluate notes and other information from electronic health records for the sampled patients to determine the "gold" or reference standard classification. Next, the measurement performance characteristic of interest (e.g., PPV) is computed for all patients sampled from all prior or current samples, who have chart reviewer determined reference values. The point estimate and associated confidence intervals are estimated after using raking weights to re-weight the risk factor distribution of the sample to resemble the overall distribution in the whole population. We compute 1-α (frequentist) confidence bands or Bayesian credible intervals for T(y). The confidence intervals are updated with each new sequential sample of size B.

The chart review process stops once a pre-specified stopping criterion is met. For instance, we might stop the chart review if the confidence bound (or the credible interval in the Bayesian case) hits some pre-specified lower or upper stopping boundary. Usually, these boundaries are not symmetric, in the sense that if the parameter of interest falls below a threshold, this might indicate that the validation is so poor that the gold standard is applied to an entire cohort, while in the case that the parameter of interest falls above a threshold, the validation is so good that further validation is unnecessary. In other scenarios, we might not be interested in pre-specified (lower and upper) stopping bounds but rather stop the chart review once a certain length/width of the confidence bounds is reached, thus reflecting a desired level of confidence irrespective of the PPV.

The precise setting is dependent on the validation under consideration. In our simulations in the next section, the stopping boundaries often depend on the target PPV, and are therefore given individually for all experiments.

**Results**

This section presents our simulation results. We start with a description of our simulation setting, followed by a visual assessment of the four sampling strategies (random sampling, stratified random sampling 1 and 2, and Neyman's sampling) and the two types of intervals (confidence bands and Bayesian credible intervals) in a single simulated run of the chart review process for a non-linked and linked scenario, where "linked" means that the patient baseline characteristics in the data matrix X are risk factors for y, the response variable. Afterwards, we quantitatively assess all techniques in a simulation study with respect to the PPV, the strength of the link between patient baseline characteristics and response, and the distribution of the strata sample sizes. We conclude by showcasing a heuristic to predict when the validation is complete while our algorithm is running.

*Simulation setup*

We consider a dataset of n=6936 frailty score measurements in Medicare enrollees with liver disease. The measured frailty score values fall into the interval [0, 0.5]. We therefore stratify them into 5 strata, given by [0,0.1), [0.1,0.2), [0.2,0.3), [0.3,0.4), and [0.4,0.5]. These values are used as the data X, which is used in stratified random sampling and Neyman's sampling strategies.

We focus on a binary response and choose the PPV as the performance characteristic of interest. We assume that our data only include patients for whom the claims based algorithm indicated that the outcome was present (for instance, y = 1). We generate a binary reference standard classification of the outcome of interest that would be obtained after chart review.

In the Sections "Comparison of the four sampling approaches in one simulated run" and "Quantitative assessment as a function of PPV" we consider two scenarios, a non-linked scenario and a linked scenario. In the non-linked scenario, the data (frailty) is not related to the binary outcome, which is generated from a Bernoulli(p) distribution, where the success probability p is chosen as the PPV. In the linked scenario, the binary outcome is generated such that there is a different success probability per stratum in such a way that the resulting PPV matches some predefined value.

In the Section "Quantitative assessment as a function of link strength", we investigate the behavior of our algorithm for different linkage strengths between the data and the response. To achieve this, we again generate the binary outcome such that there is a different success probability per stratum while ensuring that the resulting overall PPV matches some predefined value. The choice of the success probabilities per stratum allows for one degree of freedom, their standard deviation (are the success probabilities very similar across strata or very different). We regard the standard deviation of success probabilities across strata as a measure of linkage strength, where a low standard deviation indicates a low strength.

In order to investigate the behavior of our algorithm for different strata imbalances, we also vary the size of the five strata defined over [0, 0.5] that the frailty score values fall in (Section "Quantitative assessment as a function of strata balance"). They are chosen such that for our given dataset of frailty score measurements, we obtain most samples in the first stratum (which includes 0, denoted as "left skewed"), in the last stratum (which includes 0.5, denoted as "right skewed"), or we obtain a balanced allocation (denoted as "balanced").

When running our adaptive multi-wave chart review sampling process, we employ a batch size of 100. Raking is always employed to standardize the distribution of risk factors to resemble the population for which the performance characteristic is evaluated. Confidence bands are always computed with overall error α=0.05, and Bayesian credible intervals are computed by calculating the α/2-quantile $q_{\alpha/2}$ and the (1-α/2)-quantile $q_{1-\alpha/2}$ of the posterior, likewise with α=0.05. We employ the "mad" (median absolute deviation) as well as a minimum batch size of 10 for each stratum when running our simulated chart review processes. All results presented in the tables are averages of 100 repetitions.

*Comparison of the four sampling approaches in one simulated run*

Figure 2 showcases one run of a simulated chart review process in the non-linked scenario with PPV=0.8 and the two stopping boundaries ($\tau_1=\tau_2=0.75$). The rows show the four sampling strategies (random sampling, stratified random sampling 1 and 2, and Neyman's sampling), while the columns show the two types of intervals (confidence bands and Bayesian credible intervals). The horizontal lines indicate the stopping bounds. The vertical lines indicate the point at which the stopping criteria were met, in the sense that either the lower confidence limit exceeds the lower bound of the gold standard, or the upper confidence limit falls below the upper bound of the pre-specified threshold(s).

Three observations are noteworthy. First, in the depicted run, Neyman's sampling performs best in the sense that it allows for the earliest stop, with fewest charts validated, while random sampling performs worse. Second, among the two variants of stratified random sampling, the second variant using batches drawn proportionally to the strata sizes performs better. Third, Bayesian credible intervals seem to yield tighter confidence bounds than Lai's (frequentist) confidence sequences.

Figure 3 shows a similar evaluation in the case of a linked response, where the success probabilities for all strata were chosen individually but such that an overall PPV of 0.8 was met. The stopping boundaries are again ($\tau_1=\tau_2=0.75$). Here, a different picture is observed, with stratified random sampling 2 performing best for confidence bands, and Neyman's sampling performing best for Bayesian credible intervals. As before, Bayesian credible intervals seem to yield tighter confidence bounds than frequentist confidence sequences.

Figure 4 considers a different stopping criterion, the length of the confidence interval. To be precise, we stop the validation process whenever the length of the confidence interval at the current batch falls below 0.05. The simulation scenario is the same as for Figure 3, meaning a linked response with PPV=0.8. We observe that all four approaches and both types of confidence bands perform similarly, in the sense that no method seems to have an obvious edge among the others.

*Quantitative assessment as a function of PPV*

Table 1 assesses the performance of all four approaches (random sampling, stratified random sampling 1 and 2, Neyman's sampling) and the two types of intervals (confidence bands and Bayesian credible intervals) for different values of PPV. In Table 1, the PPV is varied among {0.4,0.6,0.7,0.8} while the stopping boundaries stay fixed at $\tau_1=\tau_2=0.75$.

We assess the performance of the sampling strategies with three measures. First, we show the proportion of times that the chart review process ended due to meeting one of the stopping criteria (columns 3 and 4), where a higher proportion is better. Second, we display the proportion of runs that were stopped due to futility (columns 5 and 6). The proportion of runs stopped due to sufficient or even superior efficacy is given accordingly as one minus the futility proportion. Third, we measure the number of sampling waves that were required before the stopping criterion was met (columns 7 and 8), where fewer is better.

In Table 1, we observe that all runs manage to stop. For a PPV of 0.4, the stopping times are low due to the fact that the PPV is far away from the lower stopping boundary. As the PPV increases to 0.6 and 0.7, the validation takes longer as the decision of futility is not as straightforward any more. This behavior is expected. Here, random sampling and stratified sampling 1 perform best. When the PPV is increased to 0.8, the stopping times increase further, with random sampling and stratified sampling 2 being best in connection with Lai's confidence intervals, and random sampling and stratified sampling 1 performing best in connection with Bayesian credible intervals. Moreover, Bayesian credible intervals allow for considerably shorter stopping times than frequentist Lai's confidence bands. As expected, the proportion of runs stopped due to futility is essentially one for the runs with a PPV below the stopping threshold. Interestingly, this proportion stays at one even for a PPV of 0.7, and then switches to zero suddenly for a PPV of 0.8.

A similar comparison in which the PPV is chosen in the set {0.4,0.6,0.7,0.8} while also varying the stopping boundaries $\tau_1$ and $\tau_2$ accordingly can be found in the Appendix (Table A1).

*Quantitative assessment as a function of link strength*

We aim to assess the behavior of all four methods and both confidence intervals as a function of the linkage strength between the data and the response. The setup we use, and our definition of the linkage strength, are defined in the simulation setup. The quantitative results of this experiment are given in Table 2 for a fixed PPV of 0.8 and stopping boundaries ($\tau_1=\tau_2=0.75$).

We observe in Table 2 that the stopping time seems to increase as the linkage strength increases. This is clearly visible when considering Bayesian credible intervals. Random sampling seems to perform best overall across all considered linkage strengths, while Neyman's sampling is well-suited for larger linkage strengths. The performance of stratified random sampling 1 and 2 often lies somewhere in-between the one of random and Neyman's sampling. The proportion of runs stopped due to futility is again low for a low linkage strength, and increases to around 0.3 for higher linkage strengths. This is as expected, since for a low linkage strength (standard deviation) we anticipate the futility proportion to be zero due to the choice of the PPV=0.8 and stopping thresholds $\tau_1=\tau_2=0.75$ used in this experiment. However, when increasing the linkage strength (standard deviation), we increase a higher variability into the runs, meaning we would expect a slight increase in stopping due to futility.

*Quantitative assessment as a function of strata balance*

Finally, we aim to assess the influence of the strata balance on the performance of our algorithms. To this end, we fix the PPV at 0.8, and choose the stopping boundaries as $\tau_1=\tau_2=0.75$. However, we change the definition of the four strata that was given in the simulation setup in such a way that the number of samples falling into the four strata is either left skewed, balanced, or right skewed.

Tables 3 and 4 show the results of this experiment. Table 3 considers the non-linked scenario, showing that imbalanced strata seem to yield slightly faster stopping times than balanced ones, especially in connection with Bayesian credible intervals. In particular, Neyman's sampling seems to perform very well across the scenarios considered.

Table 4 repeats the same experiment for the linked scenario with linkage strength (standard deviation) 0.05. Here, a different picture is observed, with the balanced scenario being the easiest to validate, while the two imbalanced scenarios require longer stopping times across all methods. In particular, stratified random sampling 1 and Neyman's sampling seem to perform very well across all scenarios. As before, the use of Bayesian credible intervals yields much faster stopping times than the frequentist counterpart.

In both Table 3 and Table 4, the proportion of runs stopped due to futility is essentially zero. This is as expected, since it is consistent with previous findings (Section "Quantitative assessment as a function of PPV") for the choice of PPV=0.8 and stopping thresholds $\tau_1=\tau_2=0.75$ which were used here as well.

*Prediction of stopping*

An important feature is the prediction of the point at which validation of a gold standard is possible. This is possible during runtime, in the sense that at any point, we aim to compute an estimate of the remaining number of batches and, equivalently, the remaining number of samples that would be needed.

To this end, at any point, we sample from a Bernoulli distribution with target probability set to the empirical mean of the observed data in y. This simulates a continued run of the chart review. Once the validation stops based on the updated confidence bands, we record the number of steps. By repeating this process, one can additionally compute confidence bands on the prediction.

Figure 5 shows the results of such a run. The top panel displays a run for the chart review process with stopping criteria ($\tau_1=0.20, \tau_2=0.25$). As can be seen, the validation is achieved on approaching batch number 60. At the same time, the bottom panel shows the prediction of the remaining number of batches until completion while the algorithm is running. The confidence band on the prediction is shaded in gray. We observe that the trend is captured correctly, though the number of remaining steps is underestimated by, on average, a factor of 2-4.

**Discussion**

This article considered the validation of a claims based algorithm using an efficient multi-wave chart sampling strategy to determine the reference standard and estimate performance characteristics. We investigate the use of four sampling strategies (random sampling, stratified random sampling 1 and 2, and Neyman's sampling), and the use of either frequentist confidence bands or Bayesian credible intervals to compute bounds on a binary response to be validated. We observe that Bayesian credible intervals proved to be tighter than its frequentist confidence band counterparts. Moreover, we observe that simple random sampling is often performing similarly to Neyman's sampling.

Using the simple observation that in a normal approximation, the size of confidence intervals shrinks to zero at the rate of $n^{-0.5}$, we demonstrated that it is possible to approximately predict the stopping time at which the validation is completed.

The main limitation of our study consists in the tuning of Neyman's algorithm. Indeed, Neyman's sampling is susceptible to various tuning parameters, such as the estimate of the standard deviation, the initial and successive batch sizes, the threshold for raking, or the choice of the minimal number of samples to spend per stratum and batch. Efficient ways of tuning these parameters for each scenario, and better understanding their influence on the results remains an important area of future work. Moreover, understanding better the dependence of the choice of the stopping criterion on the chart review process is an interesting area of further work. Overall, we conclude that both random sampling and Neyman's sampling, in connection with Bayesian credible intervals, allows for an efficient validation of a response of interest (compared to the approach which validates all charts), both in the binary and continuous case.

**Funding**

Supported by funds from the Division of Pharmacoepidemiology and Pharmacoeconomics of the Department of Medicine at Brigham and Women's Hospital and Harvard Medical School.

**Conflict of interest**

The authors have no conflict of interest to declare.

**References**

1. Platt R, Brown JS, Robb M, McClellan M, Ball R, Nguyen MD, Sherman RE (2018). N Engl J Med, 379(22):2091-2093.

2. EMA (2023). Real-world evidence framework to support EU regulatory decision-making – Report on the experience gained with regulator-led studies from September 2021 to February 2023. EMA/289699/2023.

3. FDA (2013). Guidance for Industry and FDA Staff – Best Practices for Conducting and Reporting Pharmacoepidemiologic Safety Studies Using Electronic Healthcare Data.

4. Lash TL, Olshan AF (2016). Epidemiology announces the "Validation Study" submission category. Epidemiology, 27:613-614.

5. Ehrenstein V, Petersen I, Smeeth L, Jick SS, Benchimol EI, Ludvigsson JF, Sørensen HT (2016). Helping everyone do better: a call for validation studies of routinely recorded health data. Clin Epidemiol, 8:49-51.

6. Fox MP, Lash TL, Bodnar LM (2020). Common misconceptions about validation studies. Int J Epidemiol, 49(4):1392-1396.

7. Collin LJ, MacLehose RF, Ahern TP, Nash R, Getahun D, Roblin D, Silverberg MJ, Goodman M, Lash TL (2020). Adaptive Validation Design: A Bayesian Approach to Validation Substudy Design With Prospective Data Collection. Epidemiology, 31(4):509-516.

8. Collins FS, Tabak LA (2014). Policy: NIH plans to enhance reproducibility. Nature, 505:612-613.

9. FDA (2024). Advancing Real-World Evidence Program.

10. Shepherd BE, Han K, Chen T, Bian A, Pugh S, Duda SN, Lumley T, Heerman WJ, Shaw PA (2023). Multiwave validation sampling for error-prone electronic health records. Biometrics, 79(3):2649-2663.

11. Amorim G, Tao R, Lotspeich S, Shaw PA, Lumley T, Shepherd BE (2021). Two-Phase Sampling Designs for Data Validation in Settings with Covariate Measurement Error and Continuous Outcome. J R Stat Soc Ser A, 184(4):1368-1389.

12. Lai TL (1976). On Confidence Sequences. Ann Statist, 4(2):265-280.

13. Hahn G and Wang SV (2024). chartreview. R-package version 1.0, available on CRAN.

14. Pasek J (2018). anesrake: ANES Raking Implementation. R-package version 0.80, available on CRAN.

**Appendix**

In case y is a continuous quantity, we might want to compute normal confidence intervals. This is done as follows. We denote with mean(y) and sd(y) the empirical mean and standard deviation of the observations in the vector y. In each wave, we compute the interval [mean(y)+$q_a$•sd(y),mean(y)+$q_{1-a}$•sd(y)] based on a normal approximation, where $q_a$ and $q_{1-a}$ denote the standard a and (1-a) normal quantiles, respectively. In order to repeatedly compute a normal interval, we use some error level a=$\alpha_i$ > 0 in each wave i which satisfies

$$\sum_{i=0}^{\infty} \alpha_i = \alpha$$

for some overall error α∈(0,1) to be spent.

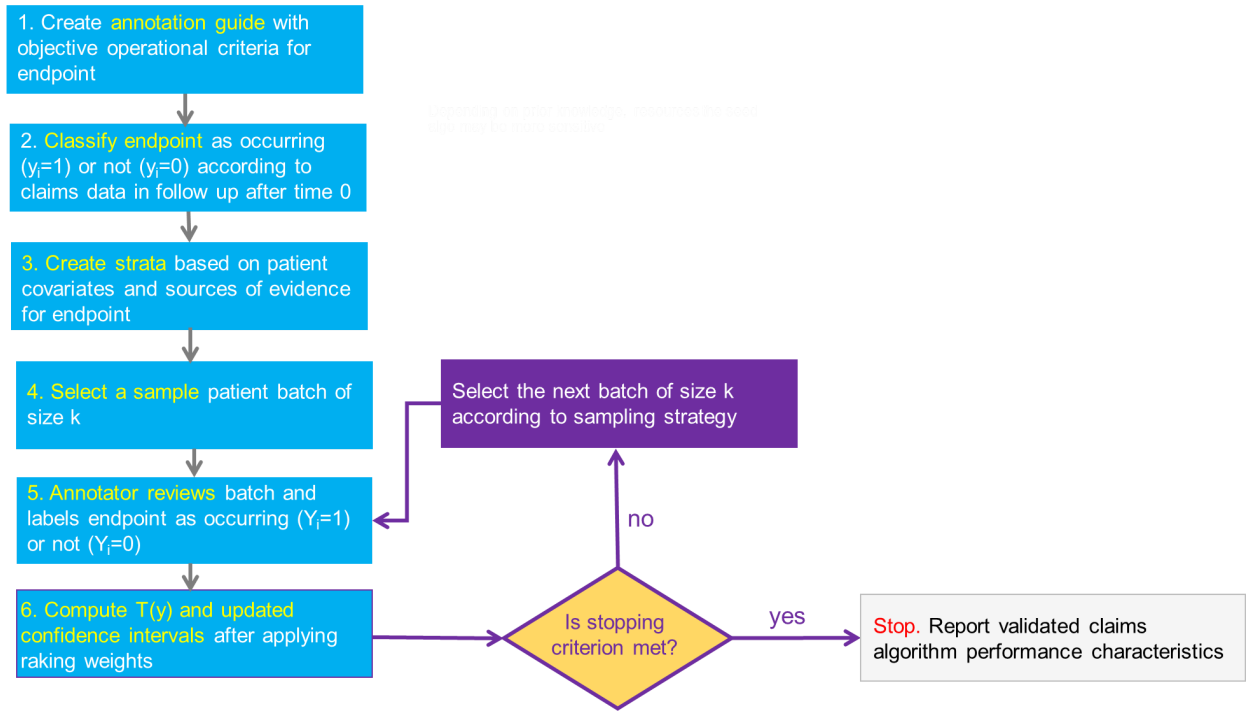

Figure 1. Flowchart of the adaptive multi-wave chart review sampling process.

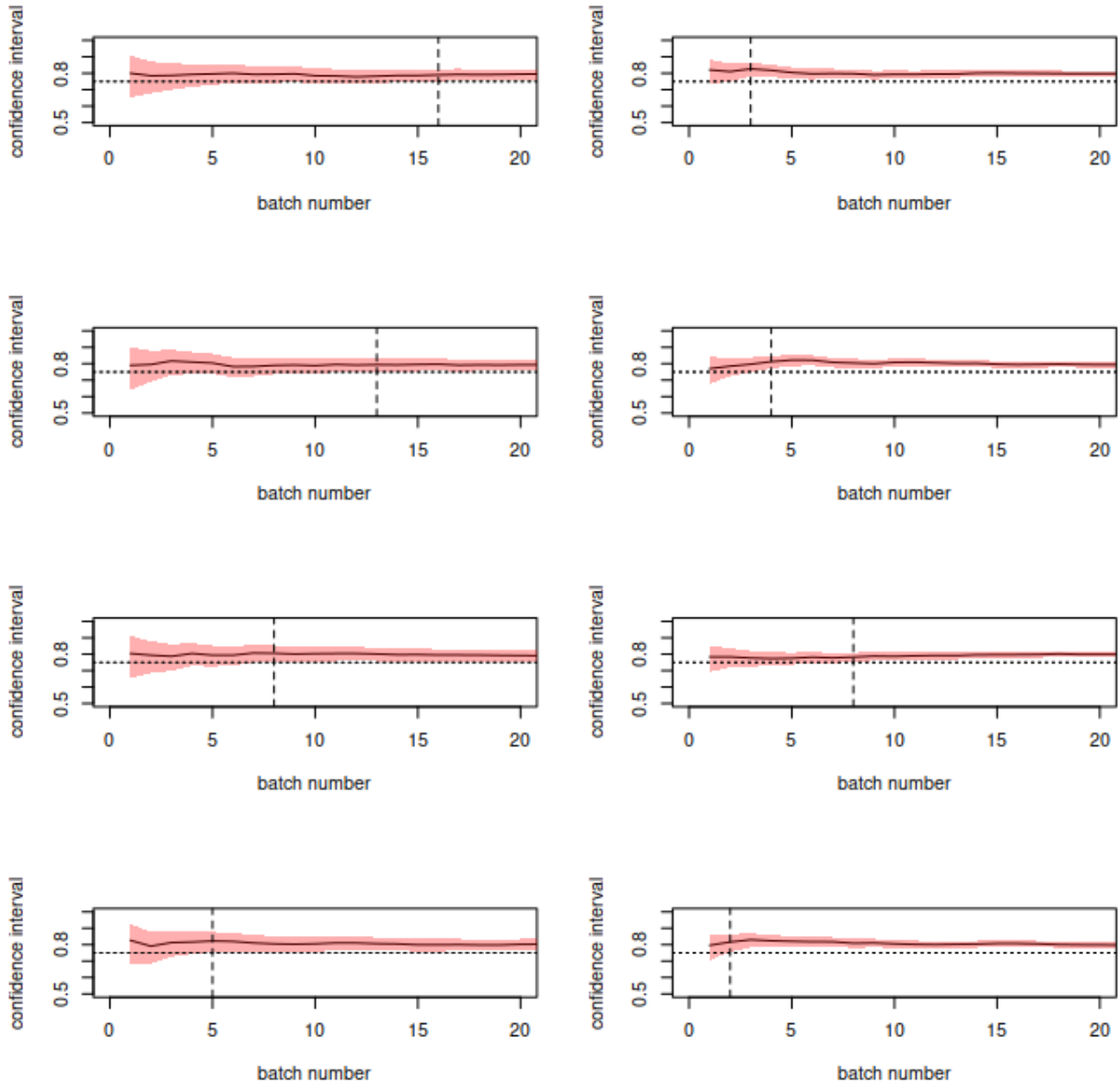

Figure 2. Estimation of the PPV in multi-wave samples when the PPV is unrelated to a risk factor for the outcome being evaluated (non-linked response with PPV=0.8 and stopping criteria $\tau_1=\tau_2=0.75$). Random sampling (first row), stratified random sampling 1 (second row), stratified random sampling 2 (third row), and Neyman's sampling (fourth row), using either Lai's confidence bands (left column) or Bayesian credible intervals (right column). The horizontal lines indicate the stopping bounds, the vertical lines indicate the stopping time.

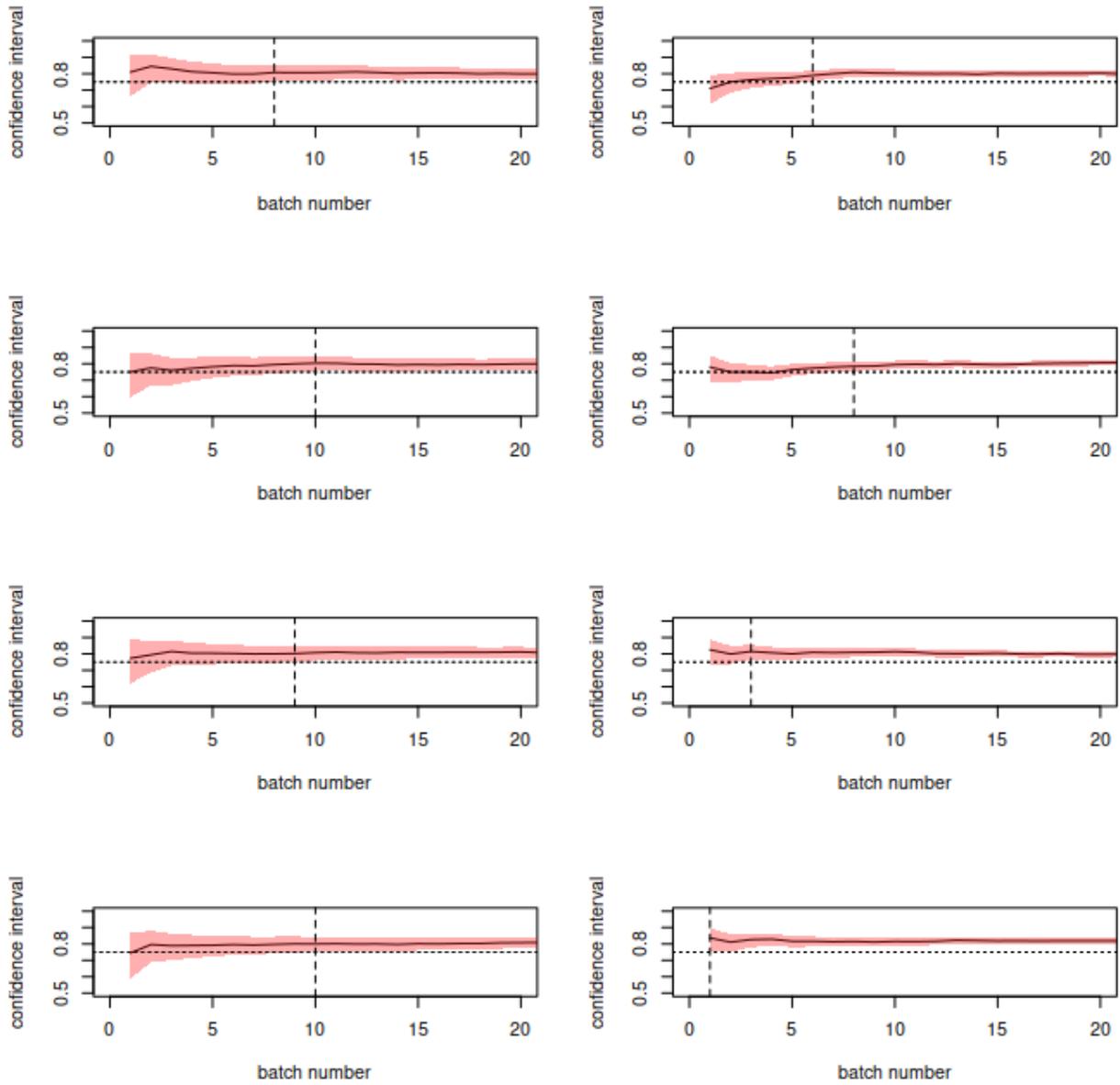

Figure 3. Estimation of the PPV in multi-wave samples when the PPV is related to a risk factor for the outcome being evaluated (linked response with PPV=0.8 and stopping criteria $\tau_1=\tau_2=0.75$). Random sampling (first row), stratified random sampling 1 (second row), stratified random sampling 2 (third row), and Neyman's sampling (fourth row), using either Lai's confidence bands (left column) or Bayesian credible intervals (right column). The horizontal lines indicate the stopping bounds, the vertical lines indicate the stopping time.

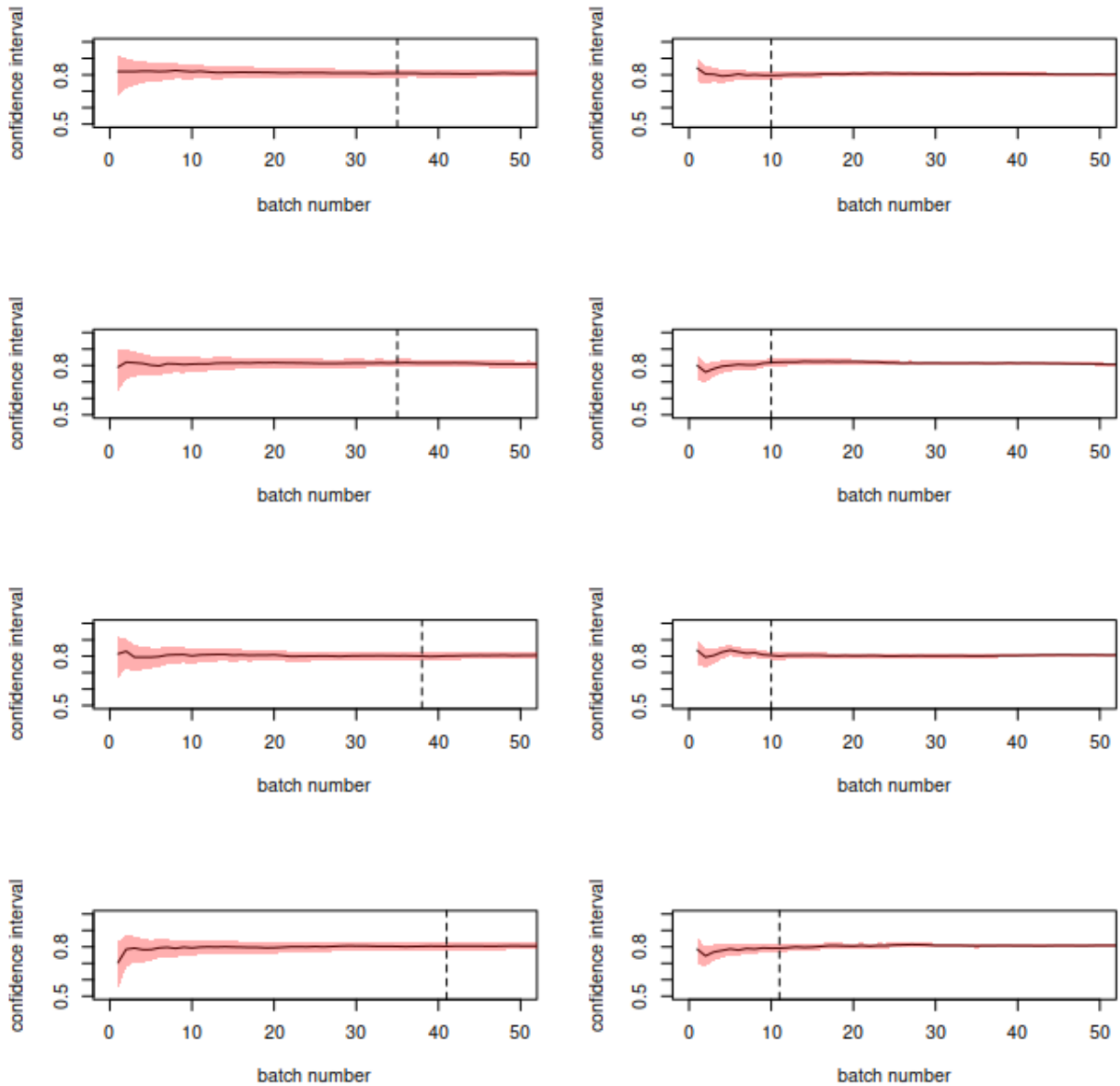

Figure 4. Estimation of the PPV in multi-wave samples when the PPV is related to a risk factor for the outcome being evaluated (linked response with PPV=0.8, stopping criterion is interval length<0.05). Random sampling (first row), stratified random sampling 1 (second row), stratified random sampling 2 (third row), and Neyman's sampling (fourth row), using either Lai's confidence bands (left column) or Bayesian credible intervals (right column). The horizontal lines indicate the stopping bounds, the vertical lines indicate the stopping time.

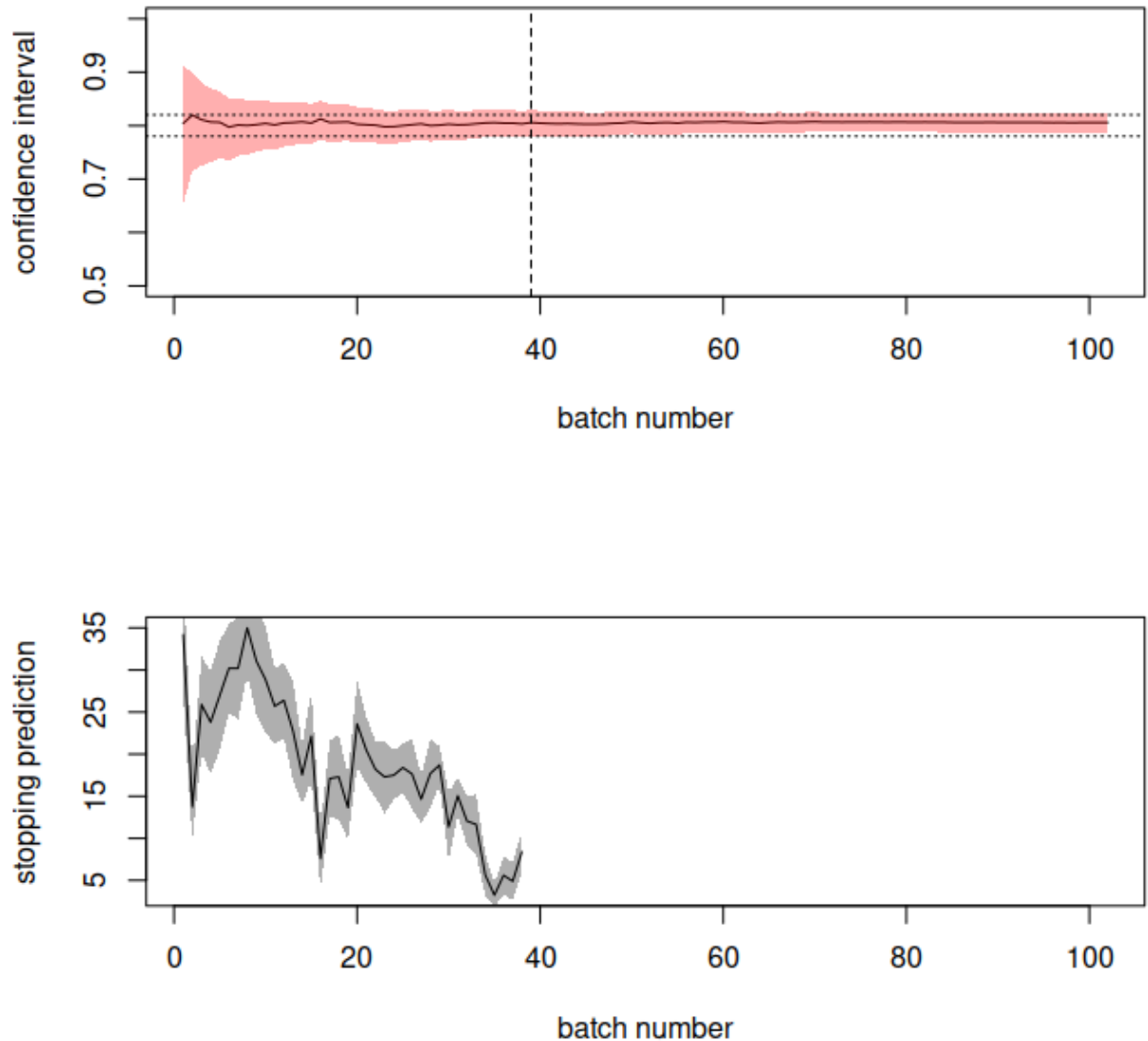

Figure 5. Prediction of the number of additional samples needed until the stopping criterion will be met. Top: Run of the multi-wave chart review process with Lai's confidence bounds and Neyman's sampling as a function of the batch number. PPV is associated with a risk factor for the outcome (linked response). Bottom: Prediction of the remaining steps until stopping as a function of the batch number including a confidence band (gray) for the estimate. Both plots have the same x-axis. The vertical lines indicate the stopping time.

| True PPV and validation interval | Method | Proportion meeting stopping criterion | | Proportion of stopping due to futility | | Number of charts reviewed until stopping | |
|---|---|---|---|---|---|---|---|
| | | Lai | Bayesian | Lai | Bayesian | Lai | Bayesian |
| PPV=0.4, ($\tau_1=\tau_2=0.75$) | random | 1.0 | 1.0 | 1.0 | 1.0 | 1.0 (1.0,1.0) | 1.0 (1.0,1.0) |
| | stratified1 | 1.0 | 1.0 | 1.0 | 1.0 | 1.0 (1.0,1.0) | 1.0 (1.0,1.0) |
| | stratified2 | 1.0 | 1.0 | 1.0 | 1.0 | 1.0 (1.0,1.0) | 1.0 (1.0,1.0) |
| | Neyman | 1.0 | 1.0 | 1.0 | 1.0 | 1.0 (1.0,1.0) | 1.0 (1.0,1.0) |
| PPV=0.6, ($\tau_1=\tau_2=0.75$) | random | 1.0 | 1.0 | 1.0 | 1.0 | 1.6 (1.4,1.7) | 1.1 (1.0,1.1) |
| | stratified1 | 1.0 | 1.0 | 1.0 | 1.0 | 1.6 (1.5,1.8) | 1.1 (1.0,1.1) |
| | stratified2 | 1.0 | 1.0 | 1.0 | 1.0 | 1.5 (1.4,1.7) | 1.2 (1.1,1.2) |
| | Neyman | 1.0 | 1.0 | 1.0 | 1.0 | 1.6 (1.5,1.8) | 1.1 (1.1,1.2) |
| PPV=0.7, ($\tau_1=\tau_2=0.75$) | random | 1.0 | 1.0 | 1.0 | 1.0 | 10.1 (8.9,11.3) | 3.8 (3.3,4.4) |
| | stratified1 | 1.0 | 1.0 | 1.0 | 1.0 | 9.4 (8.4,10.5) | 4.2 (3.5,4.8) |
| | stratified2 | 1.0 | 1.0 | 1.0 | 1.0 | 11.0 (9.8,12.3) | 3.3 (2.9,3.8) |
| | Neyman | 1.0 | 1.0 | 1.0 | 1.0 | 10.7 (9.4,11.9) | 3.6 (3.0,4.2) |
| PPV=0.8, ($\tau_1=\tau_2=0.75$) | random | 1.0 | 1.0 | 0.0 | 0.0 | 8.9 (7.9,9.8) | 3.5 (3.0,3.9) |
| | stratified1 | 1.0 | 1.0 | 0.0 | 0.0 | 9.5 (8.6,10.4) | 3.9 (3.4,4.4) |
| | stratified2 | 1.0 | 1.0 | 0.0 | 0.0 | 8.9 (7.9,9.9) | 3.5 (3.1,4.0) |
| | Neyman | 1.0 | 1.0 | 0.0 | 0.0 | 10.0 (8.9,11.1) | 3.6 (3.1,4.0) |

Table 1. Non-linked scenario. Different PPV and fixed stopping criterion. Strata defined as in the simulation setup. The stopping time is defined as the number of batches before the stopping criterion is met. Confidence intervals for the stopping times in brackets.

| SD | Method | Proportion meeting stopping criterion | | Proportion of stopping due to futility | | Number of charts reviewed until stopping | |
|---|---|---|---|---|---|---|---|
| | | Lai | Bayesian | Lai | Bayesian | Lai | Bayesian |
| 0.05 | random | 1.0 | 1.0 | 0.0 | 0.0 | 12.2 (10.1,14.3) | 5.1 (3.7,6.5) |
| | stratified1 | 1.0 | 1.0 | 0.0 | 0.0 | 11.9 (9.9,14.0) | 4.1 (3.5,4.7) |
| | stratified2 | 1.0 | 1.0 | 0.0 | 0.0 | 11.9 (9.7,14.1) | 5.8 (3.6,8.0) |
| | Neyman | 1.0 | 1.0 | 0.0 | 0.0 | 13.3 (10.6,15.9) | 4.0 (3.3,4.6) |
| 0.10 | random | 0.8 | 0.9 | 0.0 | 0.1 | 11.3 (8.6,14.1) | 8.1 (5.5,10.7) |
| | stratified1 | 0.9 | 1.0 | 0.0 | 0.0 | 11.3 (8.8,13.8) | 4.9 (2.9,6.8) |
| | stratified2 | 0.8 | 0.9 | 0.1 | 0.1 | 12.6 (9.2,16.0) | 7.7 (5.1,10.3) |
| | Neyman | 0.8 | 0.9 | 0.0 | 0.1 | 11.9 (9.0,14.9) | 6.7 (4.1,9.4) |
| 0.15 | random | 0.9 | 1.0 | 0.3 | 0.3 | 12.5 (9.6,15.3) | 6.4 (4.2,8.5) |
| | stratified1 | 0.9 | 1.0 | 0.3 | 0.0 | 20.6 (16.7,24.4) | 4.8 (3.5,6.1) |
| | stratified2 | 0.9 | 1.0 | 0.3 | 0.3 | 13.2 (10.0,16.3) | 6.7 (4.2,9.1) |
| | Neyman | 0.9 | 1.0 | 0.3 | 0.3 | 17.3 (13.5,21.0) | 9.2 (6.3,12.1) |
| 0.20 | random | 0.9 | 0.9 | 0.3 | 0.3 | 10.7 (7.8,13.7) | 5.4 (3.4,7.3) |
| | stratified1 | 1.0 | 1.0 | 0.2 | 0.0 | 18.7 (14.9,22.5) | 4.7 (3.4,6.0) |
| | stratified2 | 0.9 | 0.9 | 0.3 | 0.3 | 9.7 (6.9,12.5) | 6.2 (4.1,8.4) |
| | Neyman | 0.9 | 1.0 | 0.3 | 0.3 | 12.6 (9.0,16.1) | 8.3 (5.9,10.8) |

Table 2. Variable linkage strength at fixed PPV=0.8 and stopping criterion $\tau_1=\tau_2=0.75$. The stopping time is defined as the number of batches before the stopping criterion is met. Confidence intervals for the stopping times in brackets.

| Strata sizes | Method | Proportion meeting stopping criterion | | Proportion of stopping due to futility | | Number of charts reviewed until stopping | |
| --- | --- | --- | --- | --- | --- | --- | --- |
| | | Lai | Bayesian | Lai | Bayesian | Lai | Bayesian |
| left skewed | random | 1.0 | 1.0 | 0.0 | 0.0 | 9.6 (8.5,10.7) | 3.9 (3.4,4.4) |
| | stratified1 | 1.0 | 1.0 | 0.0 | 0.0 | 10.0 (8.8,11.1) | 3.6 (3.2,4.1) |
| | stratified2 | 1.0 | 1.0 | 0.0 | 0.0 | 9.9 (8.8,11.0) | 3.9 (3.3,4.5) |
| | Neyman | 1.0 | 1.0 | 0.0 | 0.0 | 11.5 (10.1,12.8) | 3.3 (2.9,3.8) |
| balanced | random | 1.0 | 1.0 | 0.0 | 0.0 | 9.0 (7.9,10.1) | 3.2 (2.8,3.7) |
| | stratified1 | 1.0 | 1.0 | 0.0 | 0.0 | 9.7 (8.7,10.6) | 3.5 (3.0,3.9) |
| | stratified2 | 1.0 | 1.0 | 0.0 | 0.0 | 10.3 (9.3,11.4) | 3.5 (3.0,4.0) |
| | Neyman | 1.0 | 1.0 | 0.0 | 0.0 | 9.4 (8.5,10.4) | 4.0 (3.4,4.6) |
| right skewed | random | 1.0 | 1.0 | 0.0 | 0.0 | 9.4 (8.5,10.4) | 3.5 (3.0,3.9) |
| | stratified1 | 1.0 | 1.0 | 0.0 | 0.0 | 11.0 (9.8,12.2) | 4.0 (3.4,4.6) |
| | stratified2 | 1.0 | 1.0 | 0.0 | 0.0 | 10.0 (8.8,11.2) | 3.7 (3.2,4.1) |
| | Neyman | 1.0 | 1.0 | 0.0 | 0.0 | 10.2 (9.0,11.4) | 3.9 (3.3,4.5) |

Table 3. Non-linked scenario. Strata balance for PPV=0.8 and stopping criterion $\tau_1=\tau_2=0.75$. Strata sizes which are left skewed (top), balanced (middle), and right skewed (bottom). The stopping time is defined as the number of batches before the stopping criterion is met. Confidence intervals for the stopping times in brackets.

| Strata sizes | Method | Proportion meeting stopping criterion | | Proportion of stopping due to futility | | Number of charts reviewed until stopping | |
| --- | --- | --- | --- | --- | --- | --- | --- |
| | | Lai | Bayesian | Lai | Bayesian | Lai | Bayesian |
| left skewed | random | 0.9 | 1.0 | 0.0 | 0.0 | 12.4 (10.0,14.8) | 5.8 (4.2,7.4) |
| | stratified1 | 0.9 | 1.0 | 0.0 | 0.0 | 13.5 (10.9,16.0) | 5.1 (3.6,6.5) |
| | stratified2 | 0.9 | 1.0 | 0.0 | 0.0 | 13.6 (10.8,16.4) | 5.7 (3.9,7.5) |
| | Neyman | 0.9 | 1.0 | 0.0 | 0.0 | 13.0 (10.1,15.9) | 5.9 (4.0,7.7) |
| balanced | random | 1.0 | 1.0 | 0.0 | 0.0 | 12.4 (10.9,14.0) | 3.6 (3.0,4.3) |
| | stratified1 | 1.0 | 1.0 | 0.0 | 0.0 | 9.8 (8.8,10.8) | 3.7 (3.1,4.3) |
| | stratified2 | 1.0 | 1.0 | 0.0 | 0.0 | 12.6 (11.1,14.2) | 4.4 (3.6,5.3) |
| | Neyman | 1.0 | 1.0 | 0.0 | 0.0 | 11.4 (9.9,13.0) | 3.5 (2.9,4.0) |
| right skewed | random | 1.0 | 1.0 | 0.0 | 0.0 | 12.2 (9.6,14.9) | 4.9 (3.7,6.0) |
| | stratified1 | 1.0 | 1.0 | 0.0 | 0.0 | 11.5 (9.6,13.4) | 4.0 (3.3,4.7) |
| | stratified2 | 0.9 | 1.0 | 0.0 | 0.0 | 12.3 (10.0,14.6) | 6.1 (4.2,7.9) |
| | Neyman | 0.9 | 1.0 | 0.0 | 0.0 | 11.3 (9.2,13.4) | 4.7 (3.8,5.6) |

Table 4. Linked scenario with strength(sd)=0.05. Strata balance for PPV=0.8 and stopping criterion $\tau_1=\tau_2=0.75$. Strata sizes which are left skewed (top), balanced (middle), and right skewed (bottom). The stopping time is defined as the number of batches before the stopping criterion is met. Confidence intervals for the stopping times in brackets.

| True PPV and validation interval | Method | Proportion meeting stopping criterion | | Proportion of stopping due to futility | | Number of charts reviewed until stopping | |
|---|---|---|---|---|---|---|---|
| | | Lai | Bayesian | Lai | Bayesian | Lai | Bayesian |
| PPV=0.4, ($\tau_1$=0.35,$\tau_2$=0.45) | random | 1.0 | 1.0 | 0.4 | 0.5 | 6.8 (6.2,7.4) | 2.1 (1.9,2.3) |
| | stratified1 | 1.0 | 1.0 | 0.5 | 0.6 | 7.1 (6.5,7.7) | 1.9 (1.7,2.0) |
| | stratified2 | 1.0 | 1.0 | 0.5 | 0.4 | 6.6 (6.0,7.2) | 1.9 (1.7,2.1) |
| | Neyman | 1.0 | 1.0 | 0.5 | 0.5 | 6.9 (6.3,7.4) | 2.0 (1.8,2.2) |
| PPV=0.6, ($\tau_1$=0.55,$\tau_2$=0.65) | random | 1.0 | 1.0 | 0.6 | 0.5 | 6.6 (6.0,7.2) | 2.1 (1.9,2.3) |
| | stratified1 | 1.0 | 1.0 | 0.4 | 0.5 | 7.7 (7.2,8.3) | 2.0 (1.9,2.2) |
| | stratified2 | 1.0 | 1.0 | 0.5 | 0.6 | 7.0 (6.4,7.7) | 1.9 (1.7,2.1) |
| | Neyman | 1.0 | 1.0 | 0.5 | 0.5 | 6.8 (6.2,7.3) | 2.2 (2.0,2.4) |
| PPV=0.7, ($\tau_1$=0.65,$\tau_2$=0.75) | random | 1.0 | 1.0 | 0.5 | 0.6 | 5.9 (5.4,6.4) | 1.8 (1.7,2.0) |
| | stratified1 | 1.0 | 1.0 | 0.5 | 0.5 | 6.5 (6.1,7.0) | 1.9 (1.7,2.0) |
| | stratified2 | 1.0 | 1.0 | 0.5 | 0.6 | 6.4 (5.9,6.9) | 1.9 (1.7,2.1) |
| | Neyman | 1.0 | 1.0 | 0.5 | 0.6 | 6.2 (5.7,6.7) | 1.6 (1.4,1.7) |
| PPV=0.8, ($\tau_1$=0.75,$\tau_2$=0.85) | random | 1.0 | 1.0 | 0.5 | 0.6 | 4.9 (4.5,5.3) | 1.4 (1.3,1.5) |
| | stratified1 | 1.0 | 1.0 | 0.6 | 0.6 | 4.7 (4.4,5.1) | 1.6 (1.4,1.7) |
| | stratified2 | 1.0 | 1.0 | 0.6 | 0.6 | 4.8 (4.5,5.2) | 1.6 (1.4,1.7) |
| | Neyman | 1.0 | 1.0 | 0.5 | 0.6 | 4.7 (4.3,5.1) | 1.5 (1.4,1.6) |

Table A1. Non-linked scenario. Different PPVs and corresponding stopping criteria. Strata defined as in the simulation setup. The stopping time is defined as the number of batches before the stopping criterion is met. Confidence intervals for the stopping times in brackets.